# Spin Ordering in LaOFeAs and Its Suppression in Superconductor LaO$_{0.89}$F$_{0.11}$FeAs Probed by Mössbauer Spectroscopy

Shinji KITAO[1,2], Yasuhiro KOBAYASHI[1,2], Satoshi HIGASHITANIGUCHI[1,2], Makina SAITO[1,2], Yoichi KAMIHARA[3], Masahiro HIRANO[3], Takaya MITSUI[2,4], Hideo HOSONO[3,5,6] and Makoto SETO[1,2,4]

[1]*Research Reactor Institute, Kyoto University, Kumatori-cho, Sennan-gun, Osaka 590-0494*

[2]*CREST, Japan Science and Technology Agency, Honcho Kawaguchi, Saitama 332-0012*

[3]*ERATO-SORST, JST, in Frontier Research Center, Tokyo Institute of Technology, Mail Box S2-13, 4259, Nagatsuka, Midori-ku, Yokohama 226-8503*

[4]*Japan Atomic Energy Agency, 1-1-1 Koto, Sayo-cho, Sayo-gun, Hyogo 679-5148*

[5]*Materials and Structures Laboratory, Tokyo Institute of Technology, Mail Box R3-1, 4259, Nagatsuka, Midori-ku, Yokohama 226-8503*

[6]*Frontier Research Center, Tokyo Institute of Technology, Mail Box S2-13, 4259, Nagatsuka, Midori-ku, Yokohama 226-8503*

The $^{57}$Fe Mössbauer spectroscopy was applied to an iron-based layered superconductor LaO$_{0.89}$F$_{0.11}$FeAs with a transition temperature of 26 K and its parent material LaOFeAs. Throughout the temperature range from 4.2 to 298 K, a singlet spectrum with no magnetic splitting was observed as a main component of each Mössbauer spectrum of the F-doped superconductor. No additional internal magnetic field was observed for the spectrum measured at 4.2 K under a magnetic field of 7 T. On the other hand, the parent LaOFeAs shows a magnetic transition at around 140 K, and this temperature is slightly lower than that of a structural phase transition from tetragonal to orthorhombic phase, which accompanies the resistivity anomaly at around 150 K.   The magnetic moment is estimated to be ~0.35 μB/Fe at 4.2 K in the orthorhombic phase, and the spin disorder remains in the magnetic ordered state even at 4.2 K.   The fact that no magnetic transition in $LaO_{0.89}F_{0.11}FeAs$ was observed even at 4.2 K under 7 T implies a strong spin fluctuation above $T_c$ or small magnetic moment in this system.   Therefore, the present results show that the F-doping effectively suppresses the magnetic and structural transitions in the parent material and the suppression leads to emergence of superconductivity in this system.

Since Kammerlingh Onnes found a superconductivity of Hg, materials with higher transition temperature ($T$c) have been found gradually until the discovery of high-$T$c superconducting cuprates.[1] Since when, although successive vigorous researches provided us even higher-$T$c superconductors, the race of finding new high-$T$c materials has paced down recently. At this moment, the discovery of a set of superconductors with iron-based layer, LaOFeP[2] and LaOFeAs,[3] was reported with surprises, and it is expected that it could lead a breakthrough to a further step of investigation of high-$T$c superconductors. The novel superconducting materials contain iron atoms, although the magnetic material was thought to destroy the superconducting state. There are only a few examples with iron-containing superconductors, such as filled skutterudites[4] and a high-pressure phase of iron ($\varepsilon$-Fe),[5] and the relationship between magnetism and superconductivity has been in discussion. Since the Fe in LaOFeAs has a distorted tetrahedral coordination binding covalently with 4 As, unlike a planar copper layers in cuprate superconductors, this type of materials could have a different mechanism in the superconductivity and there is a possibility to suggest a key to understanding the mechanism of the high-$T$c superconductivity. According to *ab initio* band calculations,[6-8] Fermi level is primarily composed of Fe 3d orbits. Therefore, direct information on the electronic state and magnetic moments on iron atom should be necessary to understand their nature. Mössbauer effect is one of the most powerful methods to study the electronic states of Fe. For filled skutterudites[9] and $\varepsilon$-Fe,[10] Mössbauer spectroscopy was so effective to show that no magnetic order was found in those systems. In this study, the $^{57}$Fe Mössbauer spectroscopy was applied to F-doped superconductor LaO$_{0.89}$F$_{0.11}$FeAs ($T$c=26 K) and its parent material LaOFeAs at the temperatures from 4.2 K to 298 K. A measurement with an external magnetic field of 7 T was also carried out for the F-doped superconductor to investigate the magnetic feature.

The experiment was done by a conventional Mössbauer spectroscopy using a $^{57}$Co source in Rh matrix with an activity of 1.85 GBq. Each spectrum was calibrated by an α-Fe foil and isomer shift was referenced to the α-Fe. The measurements at lower temperatures down to 4.2 K were performed using a liquid-He-flow cryostat. The measurement under a magnetic field was carried out by a cryostat with a superconducting magnet at 4.2 K under a magnetic field of 7 T. The direction of the magnetic field was parallel to the gamma-rays from the $^{57}$Co source. The $^{57}$Co source was inserted in the magnet cryostat where the magnetic field was cancelled. The magnetic field at the sample position was calibrated by a Hall probe.

The $LaO_{0.89}F_{0.11}FeAs$ and $LaOFeAs$ samples were synthesized by the method described in ref. 3 and checked by the X-ray diffraction as a single phase with only a negligible amount of FeAs and $LaAsO_4$ in LaOFeAs, and LaAs in $LaO_{0.89}F_{0.11}FeAs$. The electric resistivity measurement of the F-doped sample showed a midpoint of $T$c was 26 K and an onset of $T$c was 32 K. LaOFeAs did not show a superconducting transition, but had an anomaly at around 150 K. (See, Fig 2(a) in ref. 3.) Each sample with ~ 25 mg was mixed with BN and polyethylene powder and pressed to form a pelletized disk with a diameter of 10 mm.

Figure 1 shows typical Mössbauer spectra of the parent material LaOFeAs at the temperatures from 298 K to 4.2 K. The spectra of LaOFeAs were fit by a singlet pattern from 298 K down to 150 K. At 150 K a small amount of precursor exists beside the singlet spectrum. Then, the spectrum shows a drastic change at 140 K and magnetic-split spectra appear below 130 K down to 4.2 K. The spectrum at 140 K has a highly-distorted spectrum, indicating a complexity in the arrangement of magnetic moments.  Below 130 K down to 4.2 K, magnetic-split spectra show the existence of the internal magnetic field which is a measure of the magnetic moment. The magnetic-split spectra were fit by sextet with distributed magnetic moments. The internal magnetic fields are extracted from the peak position of its distribution as plotted in Fig. 2.  The observed internal magnetic field at 4.2 K was ~ 5.3 T.  The corresponding magnetic moment was estimated as ~ 0.35 $\mu_B$, where $\mu_B$ is Bohr magneton.  This value is close to the recently reported value of the ordered Fe magnetic moment of 0.36 $\mu_B$ at 8 K by a neutron scattering.[11]  This phase transition should be related to an anomaly observed in electric conductivity at around 150 K. This result also supports a recent identification of the structural phase transition at around 150 K by X-ray diffraction.[12]  The magnetic-split spectra were fit with small quadrupole splitting of ~ –0.03 mm/s.  Since this value is quite small, it cannot be recognized whether the singlet spectra have also this value or not.  The negative quadrupole splitting indicates a prolate distortion when the principal axis of the electric field gradient is taken along to the magnetic moment.  It should be noted that the observed magnetic-split spectra showed an enhancement of the 2nd and 5th lines. It suggests that the magnetic moment tends to lie perpendicular to the gamma-rays. Since polycrystalline sample easily oriented along c-axis due to highly anisotropic crystal structures and the c-axis possibly tends to lie parallel to the gamma-rays, the magnetic moment possibly lies perpendicular to the c-axis.

Figure 3 is typical obtained Mössbauer spectra of $LaO_{0.89}F_{0.11}FeAs$ at the temperatures from 298 K to 4.2 K.  The spectra of the F-doped superconductor appeared to be a singlet pattern throughout the temperatures from 298 K down to 4.2 K.  The fact that the spectra remain singlet below and above $T$c clearly shows that no magnetic order occurs both in superconducting and normal phases.  The isomer shifts of both samples for several temperatures are tabulated in Table I.  The observed isomer shifts increased with a decrease in temperature.  The isomer shifts of the F-doped superconductor and the parent LaOFeAs have almost similar behavior.  A related compound FeAs has an isomer shift of 0.47 mm/s and absolute value of quadrupole splitting of 0.55 mm/s at 295 K although it shows a helimagnetic structure with rotated magnetic moment of internal magnetic field between 4.8 and 1.6 T below the Néel temperature of 77 K.[13,14]  $Fe_2As$ is another iron-arsenic compound which has a Néel temperature of 353 K.  The iron in $Fe_2As$ has two sites, both of which showed a magnetic spectrum with internal magnetic fields of ~ 13 and 16 T, isomer shifts of ~ 0.5 and 0.7 mm/s, and quadrupole splittings of ~ −0.1 and −0.5 mm/s, respectively at 5 K.[15,16]  The observed isomer shift is similar to FeAs and $Fe_2As$, indicating the electron density on Fe atom has a similar behavior.  The internal magnetic moment found in LaOFeAs is roughly a similar value in FeAs, but smaller than $Fe_2As$.  On the other hand, the quadrupole splitting was estimated as quite small or almost zero in both F-doped and the parent LaOFeAs.  It is known that the quadrupole splitting is a sum of major contribution from valence electrons of the atom and a minor contribution from surrounding lattice ions.  A quite small quadrupole splitting implies that the d-electrons of the iron are arranged isotropically and that the imbalance due to tetrahedral configuration of As is very small.  It is noteworthy that this fact is not common for typical low-spin compounds which have tetrahedral arrangements and those would give a larger quadrupole splitting.  The spectra at lower than 150 K contain a small amount of magnetic-split component.  As the fitting of this small component is difficult, we cannot perform the exact identification.  However, there is a possibility that this small component is a precursor of the magnetic ordering as observed in LaOFeAs.   Otherwise, it may be due to the locally ordered Fe atoms which are almost same as the Fe atoms in LaOFeAs.   It is clear experimentally that the F-doping effectively suppresses the phase transition.[12]   The geometrical positions of random F atoms may cause the accidental locally ordered regions.   If not so, this magnetic component is considered to be contamination of the no-doped component in the F-doped sample.   The small magnetic impurity does not seem to destroy the superconductivity, but it may affect the magnetic measurements. The origin of the small magnetic component should be investigated further.

Figure 4 shows the obtained spectra under magnetic field of 7 T of the F-doped sample at 4.2 K.   The obtained internal magnetic field at 4.2 K was 7.03(3) T, while the calibrated external field was 7.04(1) T.   The fact that the observed internal field was just comparable to the external field supports no apparent magnetic order occurs in the F-doped superconductor.   Moreover, the fact that no magnetic order occurs even at 4.2 K with 7 T shows the spin moment is small in this system or the spin fluctuation is strong above $T_c$.

In conclusion, the iron in $LaO_{0.89}F_{0.11}FeAs$ shows no magnetic order throughout the temperatures from 298 K to 4.2 K.   However, the parent LaOFeAs shows a magnetic transition at around 140 K from singlet to a magnetic-split pattern, which may accompany a structural phase transition and arise the electric resistivity anomaly observed at around 150 K.   A Mössbauer spectrum under a magnetic field of 7 T at 4.2 K was also measured for $LaO_{0.89}F_{0.11}FeAs$ and proved that no magnetic order was found even at 4.2 K with 7 T.   These results clearly show that the magnetic and structural transitions were suppressed by F-doping in the LaOFeAs system, which may be a key to understanding the nature of their superconductivity.

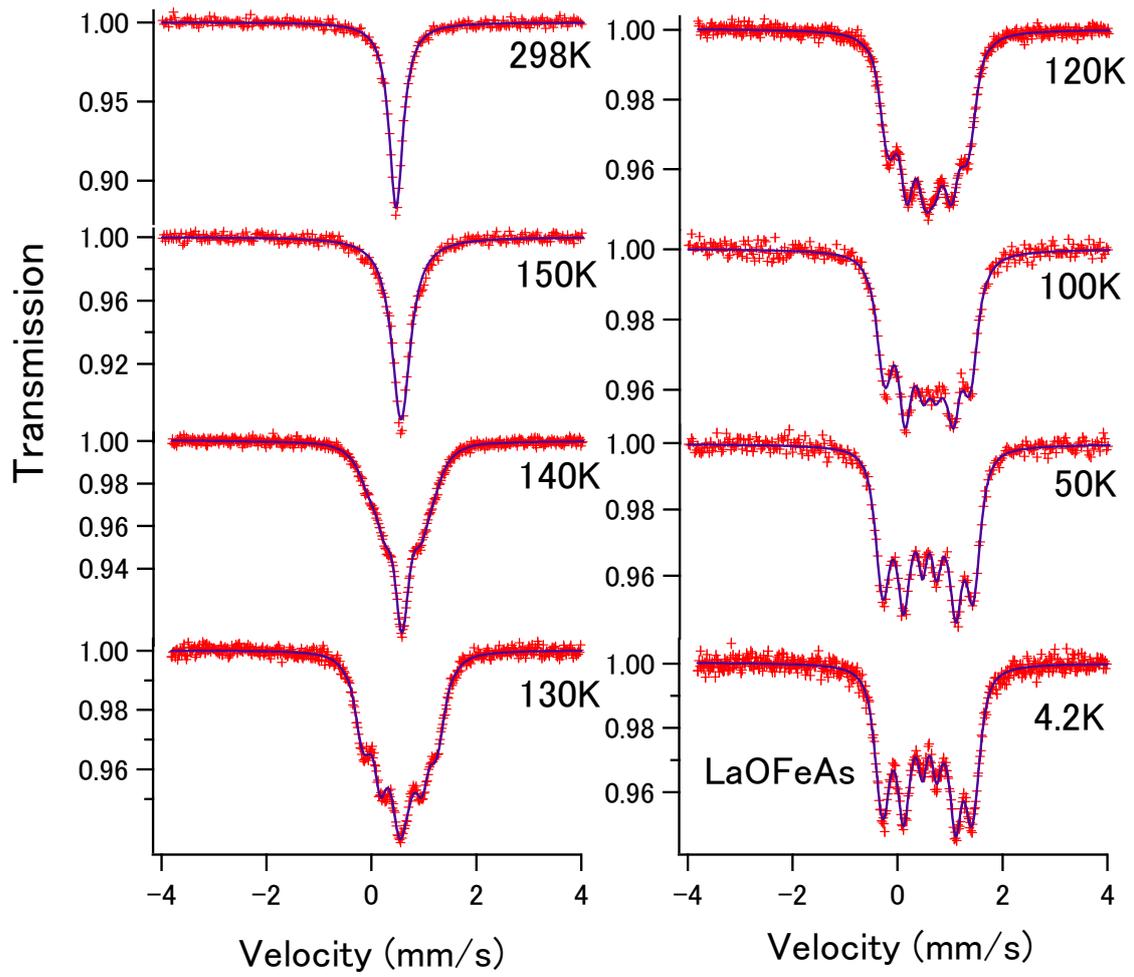

Fig. 1. Variation of Mössbauer spectra of LaOFeAs with temperatures.

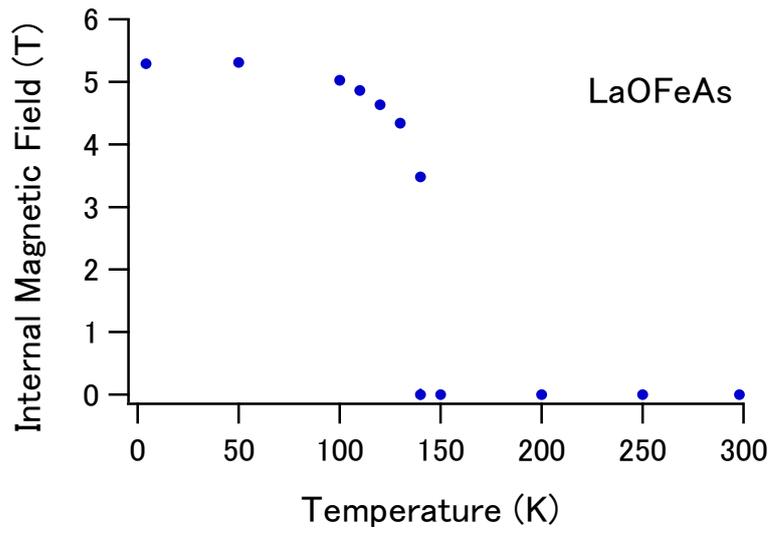

Fig. 2. Internal magnetic fields extracted from the magnetic splitting appearing in the spectra of LaOFeAs as a function of temperature.  For 140 K, two points are plotted, as it has also a singlet component.

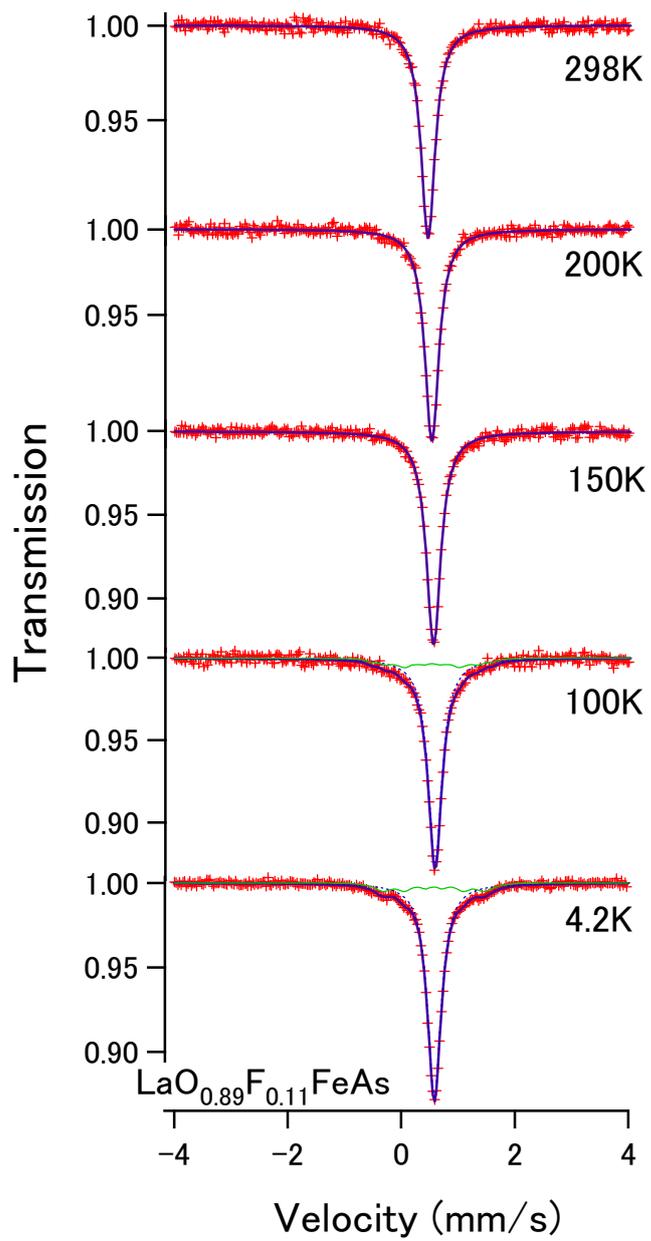

Fig. 3. Mössbauer spectra of LaO$_{0.89}$F$_{0.11}$FeAs as a function of temperature.

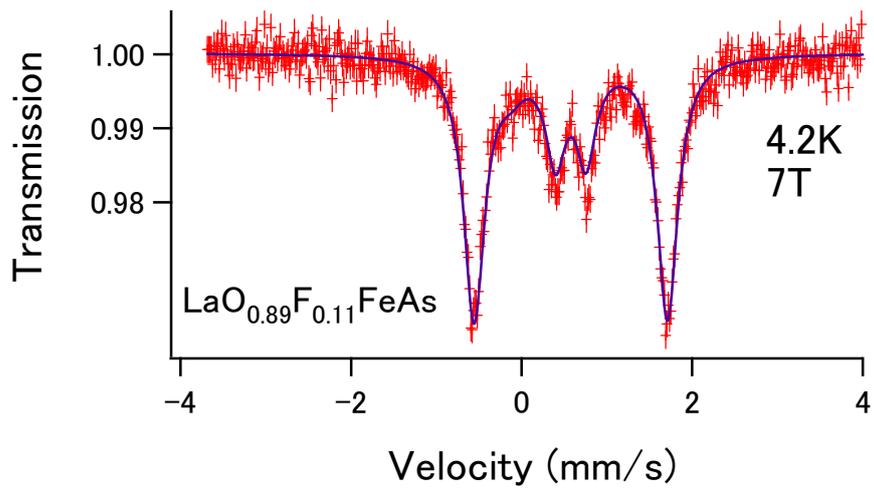

Fig. 4. Mössbauer spectrum of LaO$_{0.89}$F$_{0.11}$FeAs under a magnetic field of 7 T at 4.2 K.

Table I. Isomer shifts relative to $\alpha$-Fe of LaO$_{0.89}$F$_{0.11}$FeAs and LaOFeAs depending on the temperature.   Isomer shifts of LaOFeAs are written only when the spectrum is singlet.

| Temperature [K] | Isomer shift [mm/s] | |
|---|---|---|
| | LaO$_{0.89}$F$_{0.11}$FeAs | LaOFeAs |
| 298 | 0.45(3) | 0.44(3) |
| 200 | 0.51(3) | 0.51(3) |
| 150 | 0.54(3) | 0.54(3) |
| 100 | 0.57(3) | |
| 4.2 | 0.58(3) | |